\documentclass{mn}

\usepackage{amsfonts}
\usepackage{amsmath}
\usepackage{graphicx}

\def\gsim{\;\lower4pt\hbox{${\buildrel\displaystyle >\over\sim}$}\;}
\def\lsim{\;\lower4pt\hbox{${\buildrel\displaystyle <\over\sim}$}\;}
\def\grls{\;\lower4pt\hbox{${\buildrel\displaystyle >\over <}$}\;}
\begin{document}

\title[Axisymmetric Stability Analysis]
{Axisymmetric stability criterion for two gravitationally coupled
singular isothermal discs}

\author[Y. Shen and Y.-Q. Lou]{Yue Shen$^1$ and Yu-Qing Lou$^{1,2,3}$\\
$^1$Physics Department,The Tsinghua Center for
Astrophysics, Tsinghua University, Beijing 100084,China\\
$^2$Department of Astronomy and Astrophysics, The University
of Chicago, 5640 S. Ellis Ave., Chicago, IL 60637 USA\\
$^3$National Astronomical Observatories, Chinese Academy of
Sciences, A20, Datun Road, Beijing, 100012 China. }
%April 23, 2003 Beijing
%manuscript number MD417rv
%June 23, 2003 (Monday) resubmitted
%July 27, 2003 second report from referee Chanda J. Jog.
\date{Accepted 2003 ... Received 2003 ...;
in original form 2003 ... } \maketitle

\begin{abstract}
Using the two-fluid formalism with the polytropic approximation,
we examine the axisymmetric stability criterion for a composite
system of gravitationally coupled stellar and gaseous singular
isothermal discs (SIDs). Both SIDs are taken to be infinitely thin
and are in a self-consistent steady background rotational
equilibrium with power-law surface mass densities ($\propto
r^{-1}$) and flat rotation curves. Recently, Lou \& Shen (2003)
derived exact solutions for both axisymmetric and nonaxisymmetric
stationary perturbations in such a composite SID system and
proposed the $D_s-$criterion of stability for axisymmetric
perturbations. Here for axisymmetric perturbations, we derive and
analyze the time-dependent WKBJ dispersion relation to study
stability properties. By introducing a dimensionless stellar SID
rotation parameter $D_s$, defined as the ratio of the constant
stellar rotation speed $V_s$ to the constant stellar velocity
dispersion $a_s$, one can readily determine the axisymmetric
stability $D_s-$criterion numerically by identifying a stable
range of $D_s$. Those systems which rotate too slow (collapse) or
too fast (ring fragmentation) are unstable. We found that the
stable range of $D_s^2$ depends on the mass ratio $\delta$ of the
gaseous SID to the stellar SID and on the square of the ratio $\beta$
of the stellar velocity dispersion (which mimics the sound speed)
to the gaseous isothermal sound speed. Increment of either $\delta$
or $\beta$ or both will diminish the stable range of $D_s^2$. The
WKBJ results of instabilities provide physical explanations for
the stationary configurations derived by Lou \& Shen.
It is feasible to introduce an effective $Q$ parameter for a
composite SID system. The closely relevant theoretical studies of
Elmegreen (1995), Jog (1996) and Shu et al. (2000) are discussed. A
study of composite partial SID system reveals that an axisymmetric
dark matter halo will promote stability of composite SID system
against axisymmetric disturbances. Potential applications to disc
galaxies, circumnucleus discs around nuclei of galaxies, and
protostellar discs are briefly discussed.

%The criterion for
%the system be stable against all axisymmetric disturbances is
%addressed, in terms of an effective $Q$ parameter which is
%determined by the stellar rotation constant $D_s$, defined as the
%ratio of the stellar revolution speed $V$ to the stellar velocity
%dispersion $a_s$. The stable requirement for $Q_{eff}>1$ demands
%that $D_s^2$ must fall within a specific range. This indicates
%those systems which rotate too slow or too fast are unstable. We
%hence address that the instability presented here shows good
%explanations to the stationary configurations we have derived
%previously. The SIDs systems with too small $D_s^2$ will fall into
%the collapse region, while that with too large $D_s^2$ will fall
%into the ring-fragmentation region, which are two types of
%axisymmetric instabilities. And we found this stable range of
%$D_s^2$ depends on two parameters, one is the mass ratio of the
%gaseous disc to the stellar disc $\delta$, the other is the square
%ratio of the stellar velocity dispersion (which mimics the sound
%speed) to the gaseous isothermal sound speed $\beta$. Increment of
%either $\delta$ or $\beta$ will diminish the stable range of
%$D_s^2$. Compared with the single SID case, the composite SIDs
%system is more unstable and even can not be stable at all under
%excessive parameters. While the application of partial SIDs
%reveals that a dark-matter halo can help stabilize the composite
%system against axisymmetric disturbances.

\end{abstract}


\section{Introduction}
In the context of galactic disc dynamics, there have been numerous
investigations on criteria of axisymmetric and non-axisymmetric
disc instabilities (e.g., Binney \& Tremaine 1987).
%which play important roles in star formation.
For a single disc of either gaseous or stellar content, Safronov
(1960) and Toomre (1964) originally derived a dimensionless $Q$
parameter to determine the local stability (i.e., $Q>1$) against
axisymmetric ring-like disturbances.
%This $Q$ parameter was later proved to be a monotonic indicator
%even for non-axisymmetric instabilities (Julian \& Toomre 1966;
%Binney \& Tremaine 1987).
Besides a massive dark matter halo, a more realistic disc galaxy
involves both gas and stars. It is thus sensible to consider a
composite system of one gas disc and one stellar disc under the
gravitational control of the dark matter halo. Theoretical studies
on this type of two-component disc systems were extensive in the
past (Lin \& Shu 1966, 1968; Kato 1972; Jog \& Solomon 1984a, b;
Bertin \& Romeo 1988; Romeo 1992; Elmegreen 1995; Jog 1996; Lou \&
Fan 1998b, 2000a, b). In particular, it has been attempted to
introduce proper definitions of an effective $Q$ parameter
relevant to a composite disc system for the criterion of local
axisymmetric instabilities (Elmegreen 1995; Jog 1996; Lou \& Fan
1998b). The results of such a stability analysis may provide a
basis for understanding composite disc dynamics (e.g. Lou \& Fan
2000a, b; Lou \& Shen 2003) and for estimating global star
formation rate in a disc galaxy (e.g., Kennicutt 1989; Silk 1997).

There are several reasons that lead us to once again look into
this axisymmetric stability problem. A few years ago, Shu,
Laughlin, Lizano \& Galli (2000; see also Galli et al. 2001)
studied stationary coplanar perturbation structures in an
isopedically magnetized singular isothermal disc (SID) without
using the usual WKBJ approximation. They found exact analytical
solutions for stationary (i.e., zero pattern speed) configurations
of axisymmetric and non-axisymmetric logarithmic spiral
perturbations. According to their analysis, for axisymmetric
perturbations with radial propagation, a SID with sufficiently
slow rotation speed will collapse for sufficiently large radial
perturbation scales, while a SID with sufficiently fast rotation
speed will be unstable to ring fragmentations for relatively small
radial perturbation scales (see Fig. 2 of Shu et al. 2000). For
axisymmetric perturbations, the stable regime of SID rotation
speed may be characterized by a dimensionless rotation parameter
$D$, which is the ratio of the constant SID rotation speed $V$ to
the isothermal sound speed $a$. The critical values of the lowest
and the highest $D$ for axisymmetric stability can be derived from
their marginal stability curves. Shu et al. (2000) supported their
interpretations by invoking the familiar Toomre $Q$ parameter,
implying that a local WKBJ analysis may still have some relevance
or validity to global SID perturbation solutions. In their
proposed scheme, Shu et al. noted that the two critical values
of $D$ are fairly close to $Q=1$ for neutral stability.
%And those values of $D$ either below the lower limit or above
%the upper limit should both lead to $Q<1$ for instabilities.
As they did not invoke the WKBJ approximation in deriving the
analytic perturbation solutions, the $Q-$criterion and the
$D-$criterion correspond to each other well with large
wavenumber (ring fragmentation), while the rough correspondence
of the $Q-$criterion and the $D-$criterion for small wavenumbers
(collapse) is unexpected where the WKBJ requirement is poorly met.

Regarding astrophysical applications, these basic properties of
$D-$criterion should be properly applied in contexts even at
qualitative or conceptual levels. For example, for majority of
currently observed disc galaxies, the typical rotation parameter
$D$ may be sufficiently large to avoid the collapse regime, that
is, these disc galaxies are rotationally supported. Besides other
instabilities, the ring fragmentation
instability is conceptually  pertinent to disc galaxies such as
estimates of global star formation rates (e.g., Jog \& Solomon
1984a; Kennicutt 1989; Silk 1997; Lou \& Fan 1998b, 2002a, b).
Having said these, we should add that the collapse regime of
$D-$criterion might be relevant in early stages of galactic disc
evolution. That is, slow and fast disc rotations of proto-galaxies
may be responsible for eventual bifurcations into different
morphologies of galactic systems. Furthermore, for central discs
(with radii less than a few kiloparsecs) surrounding galactic
nuclei with supermassive black holes (e.g., Lynden-Bell 1969) or
for protostellar discs around central collapsed cores (e.g., Shu
1977), this $D$ parameter may be sufficiently small to induce disc
collapses. Therefore, both the collapse and ring-fragmentation
regimes can be of considerable interest in various astrophysical
applications.

Recently, we investigated coplanar perturbation configurations in
a composite system of two gravitationally coupled stellar and
gaseous SIDs, both taken to be razor thin (Lou \& Shen 2003). In
the ideal two-fluid formalism, we derived analytical solutions for
stationary axisymmetric and logarithmic spiral configurations in a
composite SID system in terms of a dimensionless rotation parameter
$D_s$ for the stellar SID\footnote{Subscript $_s$ indicates the
relevance to the stellar SID.}. By the analogy of a single SID
(Shu et al. 2000), these stationary configurations should form
marginal stability curves (see Fig. 2 and more extensive results
of Lou \& Shen 2003) that separate the stable region from unstable '
regions for axisymmetric disturbances. Compared with the single
SID case, the stable range of $D_s$ is reduced owing to the
presence of an additional gaseous SID and can vary for different
parameters.

In order to make convincing arguments for the above analogy and
our physical interpretations, we perform here a time-dependent
local stability analysis using the WKBJ approximation for the
composite SID system (Lou \& Shen 2003). In a proper parameter
regime, we demonstrate the correctness of interpretations by Shu
et al. (2000), and in general, we clearly show the validity of
using $D_s$ parameter to demark stable and unstable regimes. To
place our analysis in relevant contexts, we also discuss how the
two effective $Q$ parameters of Elmegreen (1995) and of Jog (1996)
are related to our $D_s$ parameter when other pertinent parameters
are specified.

In Section 2, we derive the time-dependent WKBJ dispersion relation
for the composite SID system and define several useful dimensionless
parameters. In Section 3, we present the criterion for the SID system
being stable against all axisymmetric disturbances. Finally, our
results are discussed and summarized in Section 4.

%Our basic dispersion relation for composite system has been previously
%derived by Jog \& Solomon (1984a,b), but with a different assumption
%that the epicyclic frequencies of the stellar and the gaseous are
%different, which is so demanded to satisfy our two-fluid equations
%(see Lou \& Shen 2003). We search for the criterion for a composite
%SIDs system to be stable against all axisymmetric disturbances and
%to explore the real galaxy systems for several representative types.

\section{Two-fluid composite SID system}

Here, we go through the basic fluid equations and derive the local
dispersion relation in the WKBJ approximation (Lin \& Shu 1964).
For physical variables, we use either superscript or subscript $s$
for the stellar SID and $g$ for the gaseous SID. As both SIDs are
assumed to have flat rotation curves with different speeds $V_s$
and $V_g$, we express the mean angular rotation speeds of the two
SIDs in the forms of
\begin{equation}
\Omega_s=V_s/r=a_sD_s/r,\ \qquad\  \Omega_g=V_g/r=a_gD_g/r\ ,
\end{equation}
where $a_s$ and $a_g$ are the velocity dispersion of the stellar
SID and the isothermal sound speed in the gaseous SID, respectively.
Dimensionless rotation parameters $D_s$ and $D_g$ are defined as
the ratios of stellar SID rotation speed to velocity dispersion
and gaseous SID rotation speed to sound speed, respectively. The
two epicyclic frequencies are
\begin{equation}
\begin{split}
\kappa_s\equiv\{(2\Omega_s/r)[d(r^2\Omega_s)/dr]\}^{1/2}
=\sqrt{2}\Omega_s\ ,\\
\kappa_g\equiv\{(2\Omega_g/r)[d(r^2\Omega_g)/dr]\}^{1/2}
=\sqrt{2}\Omega_g\ .
\end{split}
\end{equation}
Note that $\kappa_s$ and $\kappa_g$ are different in general.
%except for the case when $a_s=a_g$.

By the polytropic assumption for the surface mass density and
the two-dimensional pressure (e.g., Binney \& Tremaine 1987),
the fluid equations for the stellar disc, in the cylindrical
coordinates $(r,\varphi,z)$ at the $z=0$ plane, are
\begin{equation}
\frac{\partial \Sigma^{s}}{\partial t} +\frac{1}{r}\frac{\partial
}{\partial r} (r\Sigma^{s}u^{s})+\frac{1}{r^{2}}\frac{\partial
}{\partial \varphi } (\Sigma^{s}j^{s})=0\ ,
\end{equation}
\begin{equation}
\frac{\partial u^{s}}{\partial t} +u^{s}\frac{\partial
u^{s}}{\partial r} +\frac{j^{s}}{r^{2}}\frac{\partial u^{s}}
{\partial \varphi }-\frac{j^{s2}}{r^{3}}
=-\frac{1}{\Sigma^{s}}\frac{\partial }{\partial
r}(a_{s}^{2}\Sigma^{s}) -\frac{\partial \phi }{\partial r}\ ,
\end{equation}
\begin{equation}
\frac{\partial j^{s}}{\partial t}+u^{s}\frac{\partial
j^{s}}{\partial r} +\frac{j^{s}}{r^{2}}\frac{\partial
j^{s}}{\partial \varphi } =-\frac{1}{\Sigma^{s}}\frac{\partial }
{\partial \varphi }(a_{s}^{2}\Sigma^{s}) -\frac{\partial \phi
}{\partial \varphi }\ ,
\end{equation}
where $\Sigma^{s}$ is the stellar surface mass density, $u^{s}$ is
the radial component of the bulk fluid velocity, $j^{s}$ is the
$z-$component of the specific angular momentum, $a_{s}$ is the
stellar velocity dispersion (or an effective ``isothermal sound
speed"), $a_{s}^{2}\Sigma^{s}$ stands for an effective
two-dimensional pressure in the polytropic approximation, and
$\phi$ is the total gravitational potential perturbation. For the
corresponding fluid equations in the gaseous disc, we simply
replace the relevant superscript or subscript $s$ with $g$. The
two sets of fluid equations are coupled by the total gravitational
potential $\phi$ through the Poisson integral
\begin{eqnarray}
%& &\!\!\!\!
%\phi (r,\varphi ,t)=
%\nonumber \\ && \quad
\phi(r,\varphi,t)=
\oint\!d\psi\!\!\int_0^{\infty}\!\!\frac{-G\Sigma (r^{\prime
},\psi ,t)r^{\prime }dr^{\prime }}{\left[ r^{\prime
2}+r^{2}-2rr^{\prime }\cos (\psi -\varphi )\right]^{1/2}}\ ,
%\nonumber
\end{eqnarray}
where $\Sigma =\Sigma^{s}+\Sigma^{g}$ is the total surface mass
density. Here the stellar and gaseous SIDs interact mainly through
the mutual gravity on large scales (Jog \& Solomon 1984a,b; Bertin
\& Romeo 1988; Romeo 1992; Elmegreen 1995; Jog 1996; Lou \& Fan
1998b, 2000a,b; Lou \& Shen 2003).

Using the above equations for the steady background rotational
equilibrium (indicated by an explicit subscript $0$) with
$u_0^s=u_0^g=0$, $\Omega_s=j_0^s/r^2$, and $\Omega_g=j_0^g/r^2$,
we obtain
\begin{equation}
\begin{split}
\Sigma_0^s=a_s^2(1+D_s^2)/[2\pi Gr(1+\delta)]\ ,\\
\Sigma_0^g=a_g^2(1+D_g^2)\delta/[2\pi Gr(1+\delta)]\ ,
\end{split}
\end{equation}
and an important relation\footnote{This implies a close relation
among the two SID rotation speeds, the stellar velocity dispersion
and the gas sound speed. In this aspect, it is different from the
usual local prescription of a composite disc system (e.g., Jog \&
Solomon 1984a, b; Bertin \& Romeo 1988; Elmegreen 1995; Jog 1996;
Lou \& Fan 1998b). Once the ratio of $a_s^2$ to $a_g^2$ is known,
the rotation speeds of the composite SID system can be determined
through that of the stellar disc, i.e. $D_s^2$, as the two rotation
speeds are coupled dynamically. In the limiting regime of $a\ll V$
and $\beta\gg 1$, $V_s$ and $V_g$ are comparable with $V_g$ slightly
larger by an amount of $\sim (a_s^2-a_g^2)/(2V_s)$. In this sense,
the assumption $\kappa_s\approx\kappa_g$ is reasonable.}
\begin{equation}\label{ad}
a_s^2(D_s^2+1)=a_g^2(D_g^2+1)\ ,
\end{equation}
where $\delta\equiv\Sigma_0^g/\Sigma_0^s$ is the SID surface mass
density ratio, $\beta\equiv a_s^2/a_g^2$ stands for the square of
the ratio of the stellar velocity dispersion to the gas sound
speed. In our analysis, we specify values $\delta$ and $\beta$ to
characterize different composite SID systems. For late-type spiral
galaxies, gas materials are less than the stellar mass with
$\delta<1$. For young proto disc galaxies, gas materials may
exceed stellar mass in general with $\delta>1$. Thus, in our
analysis and computations, both cases of $\delta<1$ and
$\delta\geq 1$ are considered. As the stellar velocity dispersion
is usually larger than the gas sound speed ($a_s^2>a_g^2$), we
then have $\beta>1$. When $\beta=1$, we have $D_s=D_g$ by
condition (\ref{ad}), which means the two SIDs may be treated as a
single SID (Lou \& Shen 2003).

For small perturbations denoted by subscripts $1$,
we obtain the following linearized equations
\begin{equation}
\frac{\partial \Sigma_{1}^{s}}{\partial
t}+\frac{1}{r}\frac{\partial } {\partial
r}(r\Sigma_{0}^{s}u_{1}^{s}) +\Omega_s \frac{\partial \Sigma
{}_{1}^{s} } {\partial \varphi
}+\frac{\Sigma_{0}^{s}}{r^{2}}\frac{\partial j_{1}^{s}} {\partial
\varphi }=0\ ,
\end{equation}
\begin{equation}
\frac{\partial u_{1}^{s}}{\partial t}+\Omega_s \frac{\partial
u_{1}^{s}} {\partial \varphi }-2\Omega_s \frac{j_{1}^{s}}{r}
=-\frac{\partial }{\partial r}
\bigg(a_{s}^{2}\frac{\Sigma_{1}^{s}}{\Sigma_{0}^{s}}+\phi_{1}\bigg)\
,
\end{equation}
\begin{equation}
\frac{\partial j_{1}^{s}}{\partial t}+r\frac{\kappa_s
^{2}}{2\Omega_s } u_{1}^{s}+\Omega_s \frac{\partial
j_{1}^{s}}{\partial \varphi } =-\frac{\partial }{\partial \varphi
}\bigg(a_{s}^{2}\frac{\Sigma_{1}^{s}} {\Sigma_{0}^{s} }+\phi
_{1}\bigg)
\end{equation}
for the stellar SID, and the corresponding equations for the
gaseous SID with subscript or superscript $s$ replaced by $g$  in
equations (9)$-$(11). The total gravitational potential
perturbation is
\begin{eqnarray}
%& &\!\!\!\!\phi _{1}(r,\varphi ,t)=
%\nonumber \\ & &\quad
\phi_1(r,\varphi,t)= \oint\!\! d\psi\!\!\int_0^{\infty}\!\!\!
\frac{-G(\Sigma_1^s +\Sigma_1^g)r^{\prime }dr^{\prime
}}{\left[r^{\prime 2}+r^{2}-2rr^{\prime }\cos (\psi -\varphi
)\right]^{1/2}}\ .
\end{eqnarray}
In the usual WKBJ approximation, we write the coplanar
axisymmetric perturbations with the periodic dependence of
$\exp(ikr+i\omega t)$ where $k$ is the radial wavenumber and
$\omega$ is the angular frequency. With this dependence in the
perturbation equations, we obtain the local WKBJ dispersion
relation for the composite SID system in the form of
\begin{equation}\label{dis}
\begin{split}
(\omega^2-\kappa_s^2&-k^2a_s^2+2\pi G|k|\Sigma_0^s)\\
&\times
(\omega^2-\kappa_g^2-k^2a_g^2+2\pi G|k|\Sigma_0^g)\\
&-(2\pi G|k|\Sigma_0^s)(2\pi G|k|\Sigma_0^g)=0\ .
\end{split}
\end{equation}
Jog \& Solomon (1984a) have previously derived the similar
dispersion relation with $V_s=V_g$ and thus $\kappa_s=\kappa_g$.
The stability analyses of Elmegreen (1995) and Jog (1996) start
with the same dispersion relation of Jog \& Solomon (1984a).
The WKBJ approach of Jog \& Solomon (1984a) is generally
applicable to any locally prescribed properties of a two-fluid disc
system. In earlier theoretical studies including that of Jog \&
Solomon (1984a), it is usually taken that the rotation speeds of
the two fluid discs are equal. Our model of two coupled SIDs here
is self consistent globally with a central singularity and with
different SID rotation speeds in general. When one applies the
local WKBJ analysis to our composite SID system, the procedure is
identical to that of Jog \& Solomon (1984a) but with different
local SID speeds $V_s$ and $V_g$ and hence different epicyclic
frequencies $\kappa_s$ and $\kappa_g$. Therefore, our dispersion
(13) appears strikingly similar to that of Jog \& Solomon (1984a)
yet with different $\kappa_s$ and $\kappa_g$. In short, if one
prescribes two different disc speeds and thus two different
epicyclic frequencies in the analysis of Jog \& Solomon (1984a),
the resulting dispersion relation should be our equation (13).

In our analysis, the rotation speeds of the stellar and the
gaseous SIDs differ in general. In particular, since $a_s^2$ is
usually higher than $a_g^2$, it follows from equations (1) and (8)
that $\Omega_s<\Omega_g$. That is, the stellar disc rotates
somewhat slower than the gaseous disc does. Such phenomena have
been observed in the solar neighborhood as well as in external
galaxies (K. C. Freeman, 2003, private communications). This is
related to the so-called {\it asymmetric drift} phenomena (e.g.,
Mihalas \& Binney 1981) and can be understood in terms of the
Jeans equations (Jeans 1919) originally derived by Maxwell from
the collisionless Boltzmann equation. In essence, random stellar
velocity dispersions produce a pressure-like effect such that mean
circular motions become slower (Binney \& Tremaine 1987). In our
treatment, this effect of stellar velocity dispersions is modelled
in the polytropic approximation (see eqns (2) and (8)). Biermann
\& Davis (1960) discussed the possibility that the mean rotation
speeds of the stars and gas in the Galaxy may be different and
reached the opposite conclusion that a gas disc should rotate
slower than a stellar disc does. However, their analysis based on
a virial theorem that excludes the effect of stellar velocity
dispersions. In other words, the stress tensor due to stellar
velocity dispersions in the Jeans equation has been ignored in
their considerations.

%conclusion seems opposite to ours
%because in their analysis they did not correctly account the
%effect of the ``stellar pressure'', which is due to the stellar
%velocity dispersion. If all the terms are included in the
%analysis, their result will turn out to be consistent with the
%present one. In physical comprehension, the pressure term
%functions as to slow the rotation speed, so it comes naturally
%that the stars rotate slower than the gas since the stellar disc
%possesses a greater ``pressure''.

\section{Axisymmetric stability analysis}

We study the axisymmetric stability problem based on the WKBJ
dispersion relation. Similar to previous analysis (Toomre 1964;
Elmegreen 1995; Jog 1996; Lou \& Fan 1998b), we also derive an
effective $Q$ parameter. This $Q_{\hbox{eff}}$ is expected to
depend on $D_s^2$ once $\delta$ and $\beta$ are known.
%The other is a numerical
%way through the stellar velocity dispersion $D_s^2$.
%We follow the basic idea of Jog \&
%Solomon (1984a,b), but with a somewhat different assumption that
%the stellar and the gaseous discs are not necessary of the same
%epicyclic frequency $\kappa$. Hence we use subscript or
%superscript $s$ and $g$ to denote the stellar and the gaseous
%discs, respectively.
%In this manner, the dispersion relation for the composite system,
%i.e. equation (17) of Jog \& Solomon (1984a), should be revised
%into
%\begin{equation}
%\begin{split}
%(\omega^2-\kappa_s^2-k^2a_s^2&+2\pi
%Gk\Sigma_0^s)(\omega^2-\kappa_g^2-k^2a_g^2+2\pi Gk\Sigma_0^g)\\
%&-(2\pi Gk\Sigma_0^s)(2\pi Gk\Sigma_0^g)=0
%\end{split}
%\end{equation}
%where $a_s$ or $a_g$ are the velocity dispersions, and
%$\Sigma_0^s$ or $\Sigma_0^g$ are surface mass densities for
%equilibrium state.

%To simplify dispersion relation (\ref{dis}),
For the convenience of analysis, we define notations
\begin{equation}\label{1}
\begin{split}
H_1\equiv\kappa_s^2+k^2a_s^2&-2\pi G|k|\Sigma_0^s\ ,\\
H_2\equiv\kappa_g^2+k^2a_g^2&-2\pi G|k|\Sigma_0^g\ ,\\
G_1\equiv 2\pi &G|k|\Sigma_0^s\ ,\\
G_2\equiv 2\pi &G|k|\Sigma_0^g\ ,\\
\end{split}
\end{equation}
to express dispersion relation (\ref{dis}) as an explicit
quadratic equation in terms of $\omega^2$, namely
\begin{equation}\label{2}
\omega^4-(H_1+H_2)\omega^2+(H_1H_2-G_1G_2)=0\ .
\end{equation}
The two roots\footnote{The determinant
$\Delta\equiv (H_1-H_2)^2+4G_1G_2\geq0$.} $\omega_{+}^2$
and $\omega_{-}^2$ of equation (\ref{2}) are
\begin{equation}
\begin{split}
&\omega_{\pm}^2(k)=
\frac{1}{2}\{(H_1+H_2)\\
&\qquad\qquad\pm[(H_1+H_2)^2-4(H_1H_2-G_1G_2)]^{1/2}\}\ .
\end{split}
\end{equation}
The $\omega_{+}^2$ root is always positive, which can be readily
proven.\footnote{This conclusion holds true if $H_1+H_2\geq0$. If
$H_1+H_2<0$, then $H_1H_2-G_1G_2<0$ and the conclusion still
holds.}
%as already proved in Jog \&Solomon (1984a).
We naturally focus on the $\omega_{-}^2$ root, namely
\begin{equation}\label{3}
\begin{split}
&\omega_{-}^2(k)=\frac{1}{2}\{(H_1+H_2)\\
&\qquad\qquad -[(H_1+H_2)^2-4(H_1H_2-G_1G_2)]^{1/2}\}\ ,
\end{split}
\end{equation}
to search for stable conditions that make the
minimum of $\omega_{-}^2(k)\ge 0$ for all $k$.

By expanding the $\omega_{-}^2(k)$ root in terms of definition
(\ref{1}), we obtain an expression involving parameters
$\kappa_s^2$, $\kappa_g^2$, $a_s^2$, $a_g^2$, $\Sigma_0^s$,
$\Sigma_0^g$, and $k$.
%It is formidable to write out all the terms to determine
%the stable condition, which requires the minimum of
%$\omega_{-}^2(k)$ to be positive. However,
For a composite SID system with power-law surface mass densities,
flat rotation curves, and the equilibrium properties of
$\kappa_s$, $\kappa_g$, $\Sigma_0^s$ and $\Sigma_0^g$,
%
%based on our previous study (Lou \& Shen, unpublished), we
%have derived a set of conditions, as
%\begin{equation}
%\begin{split}
%\kappa_s^2=2a_s^2D_s^2/r^2\ ,\\
%\kappa_g^2=2a_g^2D_g^2/r^2\ ,
%\end{split}
%\end{equation}
%\begin{equation}
%\begin{split}
%\Sigma_0^s=a_s^2(1+D_s^2)/[2\pi Gr(1+\delta)]\ ,\\
%\Sigma_0^g=a_g^2(1+D_g^2)\delta/[2\pi Gr(1+\delta)]\ ,
%\end{split}
%\end{equation}
%and an intrinsic condition
%\begin{equation}
%a_s^2(D_s^2+1)=a_g^2(D_g^2+1)\ .
%\end{equation}
%Here $\delta$ indicates the ratio of surface mass densities of
%stellar and gaseous discs. Together with another introduced
%parameter $\beta=a_s^2/a_g^2$, i.e. the square ratio of velocity
%dispersions of two SIDs (Lou \& Shen, unpublished),
equation (\ref{3}) can also be expressed in terms of four
parameters $a_s^2$, $D_s^2$, $\delta$ and $\beta$, namely
\begin{equation}\label{4}
\begin{split}
\omega_{-}^2(K)=\frac{a_s^2}{2r^2}[A_2K^2+A_1K+A_0-\wp^{1/2}]\ ,
\end{split}
\end{equation}
where $K\equiv |k|r$ is the dimensionless local radial wavenumber
and the relevant coefficients are defined by
\begin{equation}
\begin{split}
A_2&\equiv 1+1/\beta\ ,\\
A_1&\equiv -(1+D_s^2)\ ,\\
A_0&\equiv 4D_s^2+2(1-1/\beta)\ ,
\end{split}
\end{equation}
\begin{equation}
\begin{split}
\wp\equiv B_4K^4+B_3K^3+B_2K^2+B_1K+B_0\ ,
\end{split}
\end{equation}
with
\begin{equation}
\begin{split}
B_4&\equiv (1-1/\beta)^2\ ,\\
B_3&\equiv 2(1+D_s^2)(1-1/\beta)(\delta-1)/(1+\delta)\ ,\\
B_2&\equiv (D_s^2-1+2/\beta)(D_s^2+3-2/\beta)\ ,\\
B_1&\equiv 4(1+D_s^2)(1-1/\beta)(1-\delta)/(1+\delta)\ ,\\
B_0&\equiv 4(1-1/\beta)^2\ .
\end{split}
\end{equation}
As $a_s^2$ has been taken out as a common factor of
equation (\ref{4}), the most relevant part is simply
\begin{equation}\label{5}
A_2K^2+A_1K+A_0-\wp^{1/2}
\end{equation}
that involves the stellar rotation parameter $D_s^2$ and the two
ratios $\delta$ and $\beta$. We need to determine the value of
$K_{\hbox{min}}$ at which $\omega_{-}^2(K)$ takes the minimum
value, in terms of $D_s^2$. We then substitute this
$K_{\hbox{min}}$ into expression (\ref{5}) and derive the
condition for $D_s^2$ that makes this minimal
$\omega_{\hbox{min}}^2\geq 0$.

\subsection{The $D_s-$criterion in the WKBJ regime}

To demonstrate the stability properties unambiguously, and to
confirm our previous interpretations for the marginal stability
curves in a composite SID system (Lou \& Shen 2003), we first
present some numerical results that contain the same physics and
are simple enough to be understood.

According to equation (\ref{4}), $\omega_{-}^2$ is a function of
rotation parameter $D_s^2$ and wavenumber $K$.
%$\omega_{-}^2$ may take different values when $D_s^2$ or $K$ varies.
We then use $K$ as the horizontal axis and $D_s^2$ as the vertical
axis to plot contours of $\omega_{-}^2$ numerically. A specific
example of $\delta=0.2$ and $\beta=10$ is shown in Fig. 1.
The shaded regions are where $\omega_{-}^2$ takes on negative
values and the blank region is where $\omega_{-}^2$ takes on
positive values. In this manner, we clearly obtain marginal
stability curves (solid lines) along which $\omega_{-}^2=0$.
Physically, the
shaded regions of negative $\omega_{-}^2$ values are unstable,
while the blank region of positive $\omega_{-}^2$ values is
stable. From such contour plots of $\omega_{-}^2$, one can
readily determine the specific range of stellar rotation parameter
$D_s^2$ such that the composite SID system is stable against
axisymmetric disturbances. We therefore confirm our previous
interpretations for the marginal stability curves for a composite
SID system (Lou \& Shen 2003). For a comparison, Fig. 2 shows the
corresponding marginal stability curves with $\delta=0.2$ and
$\beta=10$ that we derived previously (Lou \& Shen 2003) as the
stationary axisymmetric perturbation configuration where $\alpha$
is a dimensionless effective radial wavenumber (see also Shu et
al. 2000). The apparent difference in Figs. 1 and 2, mainly in the
collapse regions, arises because the results of Fig. 2 are exact
perturbation solutions without using the WKBJ approximation.

\begin{figure}
\begin{center}
\includegraphics[scale=0.41]{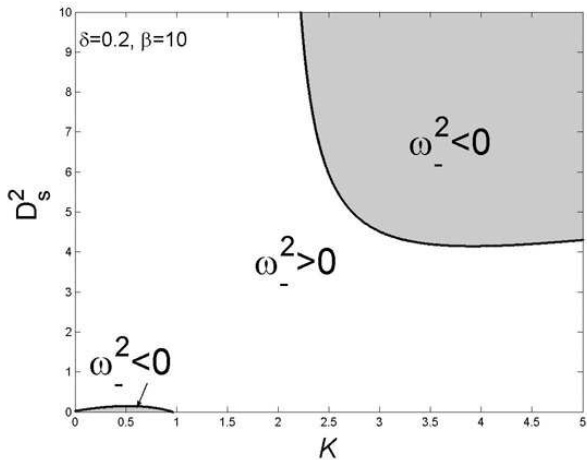}
\caption{Contours of $\omega_{-}^2$ as a function of $K$ and
$D_s^2$ with $\delta=0.2$ and $\beta=10$. The $\omega_{-}^2>0$
region is blank and the $\omega_{-}^2<0$ regions are shaded.}
\end{center}
\end{figure}

\begin{figure}
\begin{center}
\includegraphics[scale=0.43]{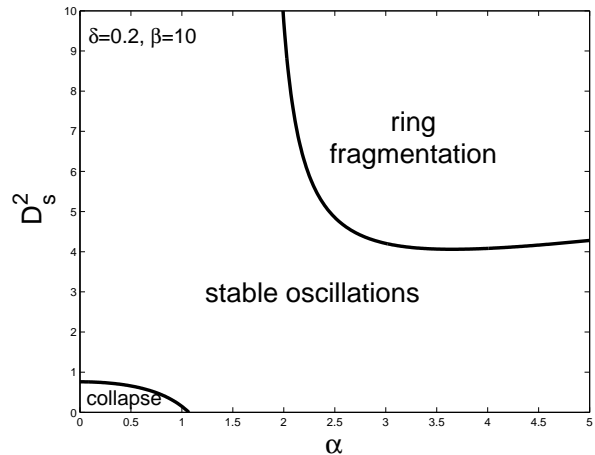}
\caption{The marginal stability curves of $D_s^2$ versus the
dimensionless effective radial wavenumber $\alpha$ for $m=0$,
$\delta=0.2$, and $\beta=10$. While $\delta$ remains small, a
larger $\beta$ lowers the ring fragmentation boundary. It is then
easier for the SID system to become unstable in the form of ring
fragmentation but with a larger $\alpha$. }
\end{center}
\end{figure}

\begin{figure}
\begin{center}
\includegraphics[scale=0.41]{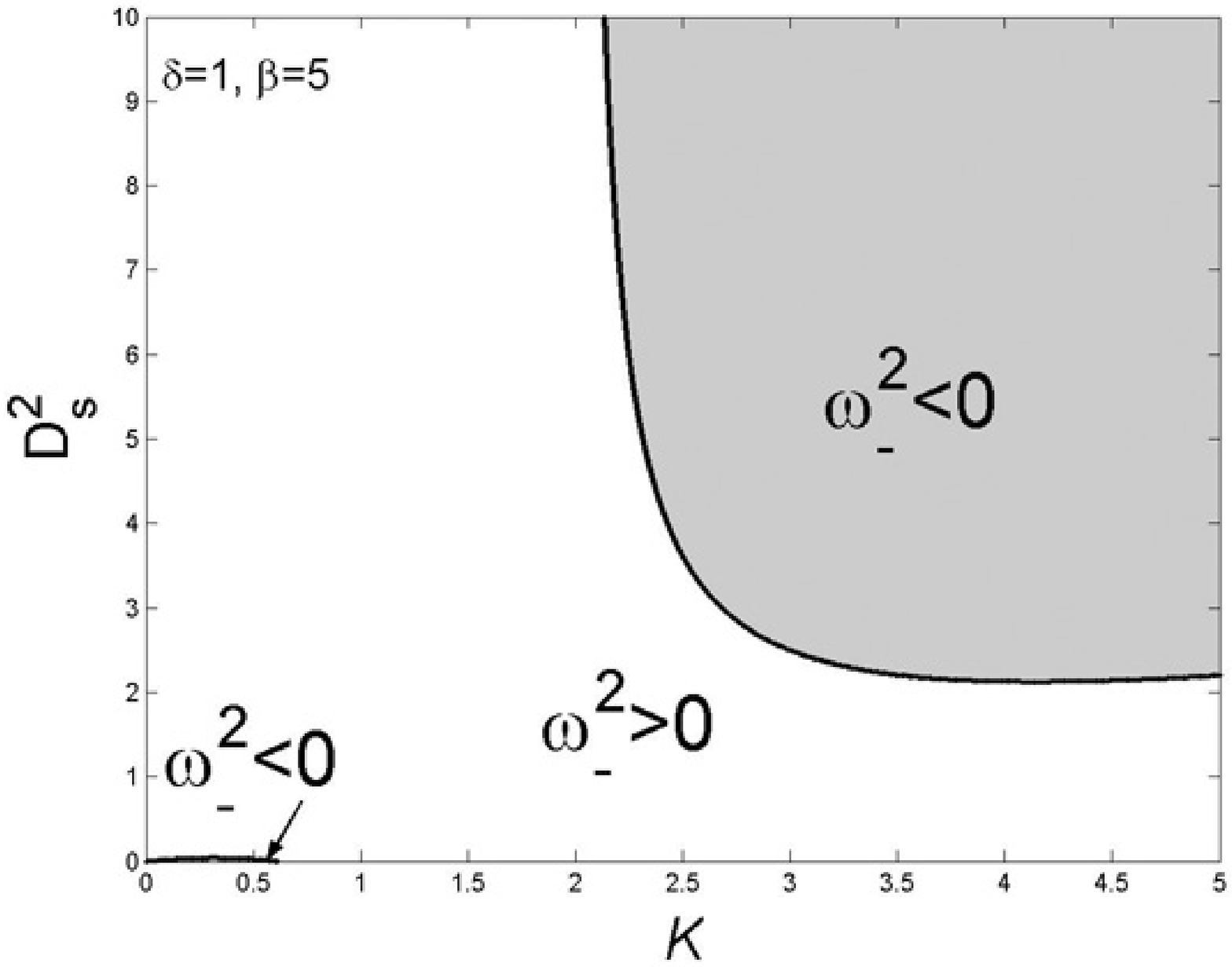}
\caption{A contour plot of $\omega_{-}^2$ as a function of $K$
and $D_s^2$ with $\delta=1$ and $\beta=5$. The $\omega_{-}^2>0$
region is blank and the $\omega_{-}^2<0$ region is shaded.}
\end{center}
%\end{figure}
%\begin{figure}[]
\begin{center}
\includegraphics[scale=0.41]{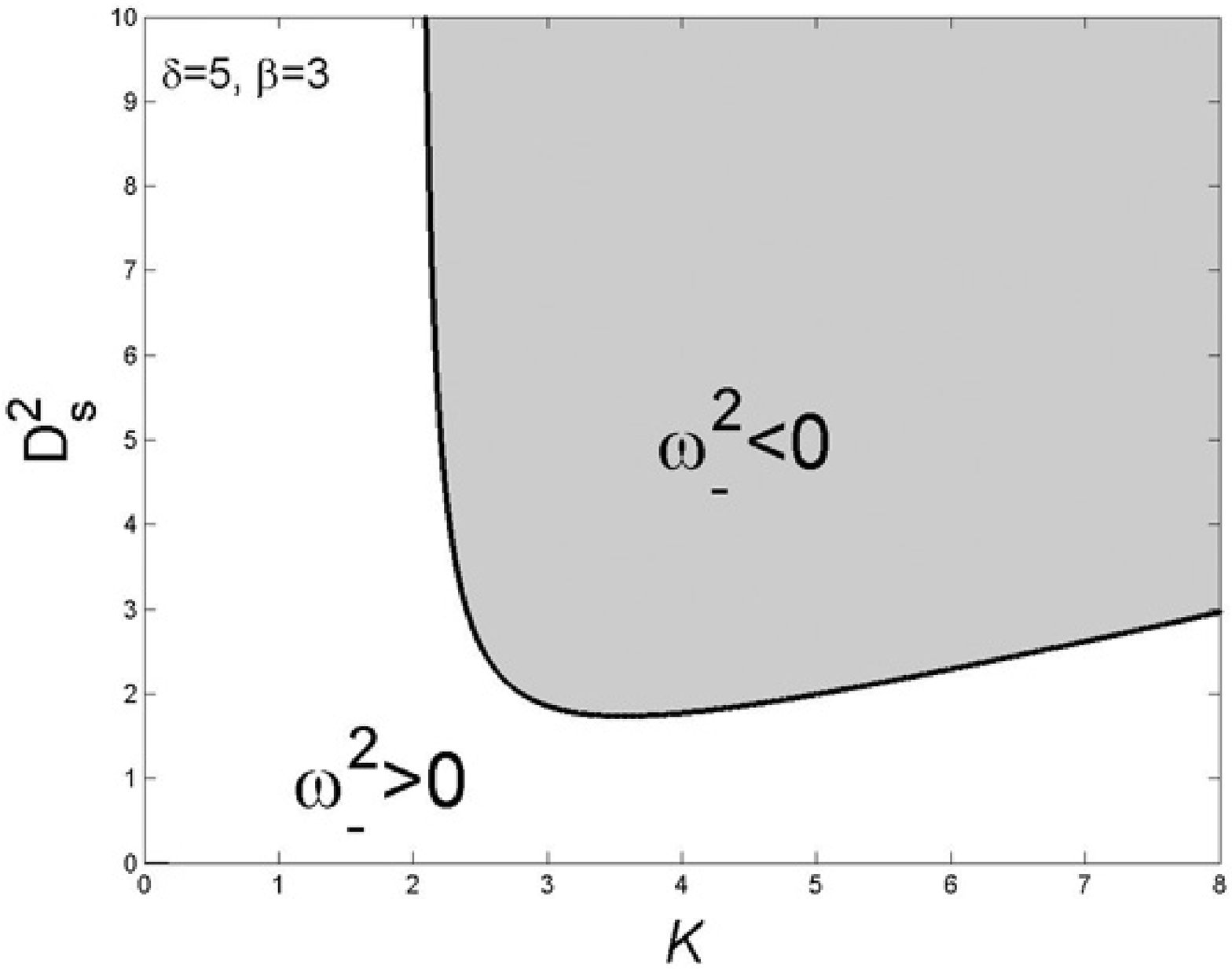}
\caption{A contour plot of $\omega_{-}^2$ as a function of $K$
and $D_s^2$ with $\delta=5$ and $\beta=3$. The $\omega_{-}^2>0$
region is blank and the $\omega_{-}^2<0$ region is shaded. }
\end{center}
\end{figure}

\begin{figure}
\begin{center}
\includegraphics[scale=0.41]{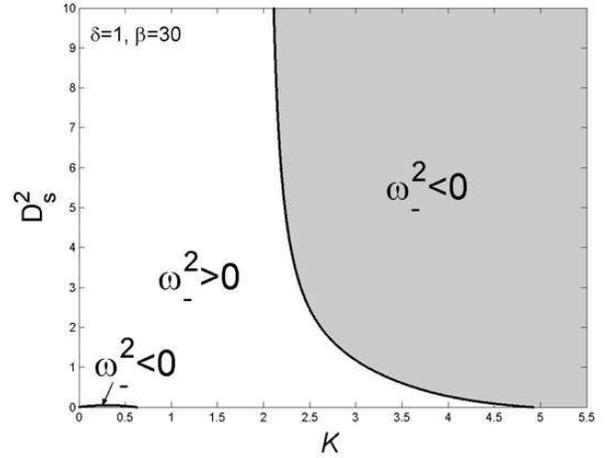}
\caption{A contour plot of $\omega_{-}^2$ as a function of $K$
and $D_s^2$ with $\delta=1$ and $\beta=30$. The $\omega_{-}^2>0$
region is blank and the $\omega_{-}^2<0$ region is shaded.}
\end{center}
\end{figure}
Now in the WKBJ approximation, the stable range of $D_s^2$ can be
read off from Fig. 1. For the case of $\delta=0.2$ and $\beta=10$,
the composite SID system is stable from $D_s^2=0.1193$ at
$K=0.5121$ to $D_s^2=4.1500$ at $K=3.9237$. In comparison with the
single SID case (see Fig. 2 of Shu et al. 2000), this stable range
of $D_s^2$ diminishes. With other sets of $\delta$ and $\beta$
parameters, we can derive similar diagrams of Fig. 2 as have been
done in our recent work, for examples, Figs. $7-10$ of Lou \& Shen
(2003). In the WKBJ approximation, we show variations with
$\delta$ and $\beta$ in Figs. 3 and 4. When $\delta$ is larger,
the system tends more likely to be unstable for small $\beta$. In
reference to Fig. 10 of Lou \& Shen (2003) with $\delta=1$ and
$\beta=30$, we plot contours of $\omega_{-}^2$ in Fig. 5 in the
WKBJ approximation. In this case, the stable range of $D_s^2$
disappears, that is, the composite SID system cannot be stable
against axisymmetric disturbances. We note again that the collapse
region with small wavenumber deviates from the exact result of
Fig. 10 of Lou \& Shen (2003), due to the obvious limitation of
the WKBJ approximation. For the unstable region of large wavenumber,
which we labelled as ring fragmentation previously (Lou \& Shen 2003),
Fig. 5 here and Fig. 10 of Lou \& Shen (2003) show good correspondence
as expected. For the purpose of comparison, Table 1
lists several sets of $\delta$ and $\beta$ and the corresponding
stable range of $D_s^2$, selected from our stationary perturbation
configurations studied earlier (see Figs. 5-10 of Lou \& Shen
2003), using both the WKBJ approximation here and the marginal
stability curves of Lou \& Shen (2003).

It is not surprising that there exist one lower limit and one upper
limit to bracket the stable range of $D_s^2$. We now provide
interpretations for the single SID case. In the usual WKBJ analyses,
the surface density $\Sigma$ is prescribed and hence independent of
the epicyclic frequency $\kappa$. In our global SID formalism, the
surface mass density and the epicyclic frequency are coupled through
the rotation parameter $D$. Both $\Sigma$ and $\kappa$ decrease with
decreasing $D$ but with different rates.
Specifically, the surface mass density $\Sigma$ is proportional
to $1+D^2$ (see eq. (7)) while $\kappa$ is proportional to $D$
(see eq. (2)). For either large or small enough $D$, the ratio
of $\kappa$ to $\Sigma$ may become less than the critical value
(sound speed is constant by the isothermal assumption), that is,
the Q parameter becomes less than unity for instabilities.

We stress that ring fragmentation instabilities occurring at
large wavenumbers are familiar in the usual local WKBJ analysis
(Safronov 1960; Toomre 1964). While collapse instabilities at
small wavenumbers are known recently for the SID system (Shu
et al. 2000; Lou 2002; Lou \& Shen 2003). The physical nature
of the new ``collapse'' regime is fundamentally due to the Jeans
instability when the radial perturbation scale becomes sufficiently
large.\footnote{One might wonder how a large-scale disc Jeans
instability can be revealed in a local WKBJ analysis although
somewhat crudely. Perhaps, the best analogy is the WKBJ dispersion
relation for spiral density waves in a thin rotating disc (Lin \&
Shu 1966, 1968). This WKBJ dispersion relation is quadratic in the
radial wavenumber $|k|$ and can be solved for a larger $|k|$ and a
smaller $|k|$ corresponding to short- and long-wave branches
respectively (e.g., Binney \& Tremaine 1987). The former is
naturally consistent with the WKBJ approximation, while the
latter, inconsistent with the WKBJ approximation, turns out to
be a necessary key ingredient in the swing process (Goldreich
\& Lynden-Bell 1965; Toomre 1981; Fan \& Lou 1997) as revealed
by analytical and numerical analyses. In the end, one needs to
perform global analyses to justify the WKBJ hint for
instabilities of relatively large scales.  }
By the conservation of angular momentum, SID rotation plays
a stabilizing role to modify the onset of Jeans collapse (e.g.,
Chandrasekhar 1961). Hence, there appears a critical $D_s^2$ below
which exists the collapse regime. Of course, too large a $D_s^2$
causes the other type of ring fragmentation instabilities to set in.
For most disc galaxies currently observed, the rotation parameter
$D_s^2$ is typically large enough to avoid such collapses.

%One might wonder how a large-scale disc Jeans instability can
%be revealed in a local WKBJ analysis although somewhat crudely.
%Perhaps, the best analogy is the WKBJ dispersion relation for
%spiral density waves in a thin rotating disc (Lin \& Shu 1966,
%1968). This WKBJ dispersion relation is quadratic in the radial
%wavenumber $|k|$ and can be solved for a larger $|k|$ and a
%smaller $|k|$ corresponding to short- and long-wave branches
%respectively (e.g., Binney \& Tremaine 1987). The former is
%naturally consistent with the WKBJ approximation, while the
%latter, inconsistent with the WKBJ approximation, turns out to
%be a necessary key ingredient in the swing process (Goldreich
%\& Lynden-Bell 1965; Toomre 1981; Fan \& Lou 1997) as revealed
%by analytical and numerical analyses. In the end, one needs to
%perform global analyses to justify the WKBJ hint for
%instabilities of relatively large scales.

Elsewhere, the collapse regime studied here may have relevant
astrophysical applications, e.g., in circumnuclear discs around
nuclei of galaxies and protostellar discs around protostellar
cores etc., where $D_s^2$ may be small enough (either a high
velocity dispersion $a_s$ or a low rotation speed $V_s$) to
initiate collapses during certain phases of disc system
evolution. In fact, for a low rotation speed $V_s$, we
note from the perspective of evolution, that those proto disc
galaxies of low rotation speeds in earlier epochs, will Jeans
collapse and lead to other morphologies, while those proto disc
galaxies of sufficiently fast rotation speeds in earlier epochs
will gradually evolve into disc galaxies we observe today.
Meanwhile, ring fragmentation instabilities are thought to be
relevant to global star formation in a disc galaxy -- another
important aspect of galactic evolution.

The addition of a gaseous SID to a stellar SID will decrease the
overall stability against axisymmetric disturbances at any
wavelengths in general, that is, the stable range of
$D_s^2$ shrinks. However, for the two types of instabilities,
namely, the collapse instability at small wavenumbers and the
ring-fragmentation instability at relatively large wavenumbers,
the gravitational coupling of the two SIDs play different roles.
For instance, a composite SID system tends more likely to fall
into ring fragmentations but less likely to collapse. For
either larger $\delta$ or larger $\beta$, this tendency
becomes more apparent.

This trend can be understood physically. For the
ring-fragmentation regime, suppose the stellar SID by
itself alone is initially stable and a gaseous SID
component is added to the stellar SID system in a
dynamically consistent manner. The gaseous SID may be either
stable or unstable by itself alone. By itself, the gaseous SID
tends to be more unstable if the sound speed $a_g$ becomes smaller
or $\Sigma_0^g$ becomes larger according to the definition of
the $Q$ parameter (valid for the ring fragmentation instability),
which means larger $\beta$ or larger $\delta$, respectively. Thus,
with either $\beta$ or $\delta$ or both being larger, a composite
SID system tends to be more vulnerable to ring fragmentations.
The interpretation for the collapse regime is associated with the
dynamical coupling of surface mass density $\Sigma$ and epicyclic
frequency $\kappa$ that leads to the following conclusion, namely,
a lower velocity dispersion of the gas component seems to prevent
collapse. This should be understood from condition (8): for larger
$\beta$, the rotation speed $V_g$ will exceeds $V_s$ by a larger
margin and helps to prevent an overall SID collapse.

As already noted, the WKBJ approximation becomes worse in the
quantitative sense when dealing with small wavenumbers where the
collapse regime exists. But the stability properties of a SID can
be qualitatively understood through the WKBJ analysis. The exact
perturbation analysis of our recent work (Lou \& Shen 2003) has
shown that the collapse regime diminishes with increasing values
of $\delta$ and $\beta$. Therefore, once the interpretations are
justified physically, the exact procedure of Lou \& Shen (2003)
should be adopted to identify the relevant stable range of $D_s^2$
of a composite SID. As much as Shu et al. (2000) have done for a
single SID analytically, Lou \& Shen (2003) did the same for a
composite SID system with exact solutions to the Poisson integral.
The stationary axisymmetric perturbation configurations were
derived to determine the marginal stability curves. These marginal
stability curves have been proven to have the same physical
interpretation as the WKBJ marginal curves obtained here.
%Hence, the exact stationary
%perturbation configurations are preferred to describe the stability
%properties of a composite SID system.
\begin{table}
\begin{center}
\begin{tabular}{cccc}
\hline\hline $\delta$ & $\beta$ & lower limit  & higher limit \\
&&of $D_s^2$&of $D_s^2$\\
\hline - & 1 & 0.1716(0.9261) & 5.8284(5.6259)\\
\hline 0.2 & 1.5 &0.1350(0.8406) &5.3974(5.2117)\\
 & 5  & 0.1207(0.7705) &4.4782(4.3558)\\
 & 10 &0.1193(0.7602)&4.1500(4.0611)\\
 & 30 & 0.1185(0.7539)  &3.8062(3.7670)\\
\hline 1&1.5&0.0596(0.6556)  &4.5945(4.4352)\\
&5 &0.0411(0.4478) &2.1275(2.0853) \\
&10&0.0401(0.4221)  & 1.0947(1.0834)\\
&30&-(-) &-(-)\\
\hline 5&3&0.0039(0.1418) &1.7397(1.6718)\\
\hline 10&5 &0.0011(-) &0.5696(0.5317)\\
\hline
\end{tabular}
\caption{The overall stable ranges of $D_s^2$ against axisymmetric
disturbances at any wavelengths for different sets of $\delta$ and
$\beta$. The values in parentheses are derived by Lou \& Shen (2003)
without the WKBJ approximation and are preferred to describe the
collapse region. The overall stable range of $D_s^2$ appears to
decrease when either $\delta$ or $\beta$ increases. The stable
range of $D_s^2$ may disappear for a case like $\delta=1$ and
$\beta=30$. When $\beta=1$, the problem is independent of $\delta$,
that is, $\delta$ can be arbitrary (Lou \& Shen 2003). }
\end{center}
\end{table}
%For convenience, we change variable $k$ into $K=kr$, hence we can
%remove the parameter $r$ from the coefficients. With this
%modification, equation (\ref{4}) turns out to be
%\begin{equation}\label{8}
%\begin{split}
%\omega^2(K)=\frac{a_s^2}{2r^2}[A^{\prime}_2K^2+A^{\prime}_1K
%+A^{\prime}_0-\wp^{\prime1/2}]\ ,
%\end{split}
%\end{equation}
%where
%\begin{equation}
%\begin{split}
%\wp^{\prime}=B^{\prime}_4K^4+B^{\prime}_3K^3+B^{\prime}_2K^2
%+B^{\prime}_1K+B^{\prime}_0\ .
%\end{split}
%\end{equation}
%Here the coefficients $A^{\prime}_2-A^{\prime}_0$ and
%$B^{\prime}_4-B^{\prime}_0$ are obtained by removing the parts
%associated with $r$ from $A_2-A_0$ and $B_4-B_0$.

\subsection{Effective $Q$ parameters}

It is natural to extend the concept of the $Q$ parameter for a
single disc to an effective $Q$ parameter for a composite disc
system for the purpose of understanding axisymmetric stability
properties. In the following, we discuss effective $Q$ parameters
for a composite SID system in reference to the works of Elmegreen
(1995) and Jog (1996). There are several points worth noting in
our formulation. First, $\kappa_s$ and $\kappa_g$ are related to
each other but are allowed to be different in general. Secondly,
the background surface mass densities are related to SID rotation
speeds through the polytropic approximation and the steady
background rotational equilibrium. For the convenience of notations,
the effective $Q$ parameters introduced by Elmegreen (1995) and by
Jog (1996) are denoted by $Q_E$ and $Q_J$, respectively.

\subsubsection{The $Q_E$ parameter of Elmegreen}

Starting from equation (18) or (22), we define an effective $Q_E$
parameter in a composite SID system following the procedure of
Elmegreen (1995) but with two different epicyclic frequencies
$\kappa_s$ and $\kappa_g$. The minimum of $\omega^2_{-}$ is given
by
\begin{equation}\label{6}
\begin{split}
\omega^2_{-min}&=\frac{a_s^2}{2r^2}[A_2K_{min}^2+A_1K_{min}+A_0
-\wp^{1/2}]\\
&=\frac{a_s^2}{2r^2}(\wp^{1/2}-A_2K_{min}^2-A_1K_{min})(Q_E^2-1)\
,
\end{split}
\end{equation}
where
\begin{equation}\label{Qeff}
Q_E^2\equiv\frac{A_0}{\wp^{1/2}-A_2K_{min}^2-A_1K_{min}}\
\end{equation}
and $\wp$ takes on the value at $K=K_{min}$. As $K_{min}$ depends
on $D_s^2$, $\delta$ and $\beta$, so does $Q_E^2$. For specified
parameters $\delta$ and $\beta$ of a composite SID system, the
parameter $Q_E^2$, giving the criterion of axisymmetric stability
(i.e. stable when $Q_E^2>1$), corresponds to the rotation
parameter $D_s^2$ of the stellar disc.
%So it is conceptually correct that
%only within a range of $D_s^2$ can the SIDs system be stable
%against axisymmetric disturbances.
One purpose of the present analysis is to establish the
correspondence between the axisymmetric stability condition and
the marginal stability curves of axisymmetric stationary
perturbation configuration derived recently by Lou \& Shen (2003)
for a composite SID system. The pertinent results for a single SID
can be found in Shu et al. (2000).

Our task now is to find the solution $K_{min}$. As
$\kappa_s\neq\kappa_g$ in general, our computation turns out to be
somewhat more complicated or involved than that of Elmegreen (1995)
who took $\kappa_s=\kappa_g$. It becomes more difficult to solve
this algebraic problem of order higher than the third analytically.
%However, we can find the critical values for $D_s^2$ with
%$Q_{eff}=1$. The critical values of $D_s^2$ can be derived
%through the following set of two simultaneous equations,
However, if one multiplies both sides of equation (17) by the
positive root $\omega_{+}^2$, a simple expression appears
\begin{equation}
{\cal W}\equiv\omega_{+}^2\omega_{-}^2=H_1H_2-G_1G_2\ .
\end{equation}
Instead of minimizing $\omega_{-}^2$, we solve for $K_{c}$ in
terms of $D_s^2$ at which ${\cal W}\equiv\omega_{+}^2\omega_{-}^2$
attains the minimum value. This $K_{c}$ must satisfy the following
equation
\begin{equation}
\begin{split}
\frac{d{\cal W}}{dK}=\frac{d(H_1H_2-G_1G_2)}{dK}=0\ .
\end{split}
\end{equation}
%\begin{equation}
%\begin{split}
%\frac{d\omega^2(K)}{dK}=0\\
%\omega^2(K)=0\ .
%\end{split}
%\end{equation}
%Together with equation (\ref{3}) we obtain
%\begin{equation}\label{set2}
%\begin{split}
%d(H_1H_2-G_1G_2)/dK=0\\
%H_1H_2-G_1G_2=0\ ,
%\end{split}
%\end{equation}
%which should be solved simultaneously. Note the second equation is
%obtain by multiplying both side of equation (\ref{3}) by the
%positive root of $\omega^2$.
By substituting $H_1$, $H_2$, $G_1$ and $G_2$ into equation (26)
and regarding $\delta$ and $\beta$ as specified parameters,
equation (26) can be reduced to a cubic equation of $K$ involving
$D_s^2$, namely,
\begin{equation}\label{K3}
\begin{split}
K^3+aK^2+bK+c=0\ ,
%\frac{1}{4}K^4+\frac{a}{3}K^3+\frac{b}{2}K^2+cK+d=0\ ,
\end{split}
\end{equation}
where
$$a=-3(D_s^2+1)(1+\beta\delta)/[4(1+\delta)]\ ,$$
$$b=D_s^2(\beta+1)+\beta-1\ ,$$
$$c=-(D_s^2+1)[\beta D_s^2(1+\delta)+\beta-1]/[2(1+\delta)]\ .$$
%$$d=D_s^2(\beta D_s^2+\beta-1)$$. It can be proved that
In most cases \footnote{When $\kappa_s=\kappa_g$, Bertin \& Romeo
(1988) give the proof that as long as $1/\beta>0.0294$ and
$\delta>0.172$, there exists only one minimum of $\omega_{-}^2$.
When there are more than one local minima for $\omega_{-}^2$, one
should choose $K_c$ for the smallest minimum of $\omega_{-}^2$.},
equation (27) has only one real solution in the form of
\begin{equation}
K_{c}=(x-q/2)^{1/3}+(-x-q/2)^{1/3}-a/3\ ,
\end{equation}
where $x=(q^2/4+p^3/27)^{1/2}$, $p=b-a^2/3$ and
$q=2a^3/27-ab/3+c$. For a single real root, it is required that
$q^2/4+p^3/27>0$ . Through numerical computations, we find that
this condition is met for most pairs of $\delta$ and $\beta$.

%Substitute this $K_{c}$ into the second equation of
%(\ref{set2}) and solve it for the critical values of
%$D_s^2$. Hence we have determined the stable range of $D_s^2$.
We then regard this $K_c$ as an estimator\footnote{While $K_c$ is
obtained as the value at which $\omega_{-}^2$ multiplying
$\omega_{+}^2$ reaches the minimum. It is valid to use $K_{c}$ to
determine whether $Q_E^2>1$ or $Q_E^2<1$ by realizing that
$\omega_{+}^2$ remains positive and the critical condition
$Q_E^2=1$ is equivalent to $\omega_{+}^2\omega_{-}^2=0$.} for
$K_{min}$ to determine $Q_E^2$.
%For a limitation case when $\delta=0$, which means
%there is only one stellar disc, then the situation
%should degenerate to the single disc case.
As already known in our analysis (Lou \& Shen 2003), when the
ratio $\beta$ is equal to 1, the properties of a composite SID
system have something in common with those of a single SID.
Moreover, the stationary axisymmetric configuration becomes the
same as a single SID. We expect that the present problem should
reduce to a single SID case when $\beta=1$.
%Meanwhile, this $\beta=1$ case is really
%a limitation for real two-SID systems.
We consider below the special case of $\beta=1$.

With $\beta=1$ in equation (\ref{K3}), $K_c$ can be determined in
terms of $D_s^2$ without involving $\delta$ at all. We then use
$K_c$ as $K_{min}$ in the definition of $Q_E^2$, namely, condition
(\ref{Qeff}). One can readily verify that $Q_E^2>1$ for
$3-2\sqrt{2}<D_s^2<3+2\sqrt{2}$; $Q_E^2<1$ for $D_s^2>3+2\sqrt{2}$
or $D_s^2<3-2\sqrt{2}$; and $Q_E^2=1$ for $D_s^2=3\pm2\sqrt{2}$.
%\begin{equation}
%\begin{split}
%\omega^2(K)=\frac{a_s^2}{r^2}[K^2-(1+D_s^2)K+2D_s^2]\ .
%\end{split}
%\end{equation}
%It is easy to find
%\begin{equation}
%\begin{split}
%\omega_{min}^2=\frac{a_s^2(1+D_s^2)^2}{4r^2}(Q_{eff}-1)\\
%at\ K_{min}=(1+D_s^2)/2\ ,
%\end{split}
%\end{equation}
%where
%\begin{equation}
%Q_{eff}=2\sqrt{2}\frac{D_s}{1+D_s^2}
%\end{equation}
The range of $D_s^2$ that makes $Q_E^2>1$ is
$0.1716<D_s^2<5.8284$, which is exactly the same as the $Q$
estimator for the single SID of Shu et al. (2000). Based on
estimates of $Q$ parameter alone, Shu et al. (2000) made physical
interpretations for their axisymmetric stationary perturbation
solution curves for a single SID. According to their
interpretation, the maximum $D^2=0.9320$ of the collapse branch
and the minimum $D^2=5.410$ of the ring-fragmentation branch (see
Fig. 2 of Shu et al. 2000) encompass the required range of $D^2$
for stability against axisymmetric disturbances. As the
formulation of Shu et al. (2000) is exact in contrast to the WKBJ
approximation, the stable range of $D^2$ slightly deviates from
that determined by the effective $Q_E$ parameter. And for the
upper bound of $D^2$, with a larger wavenumber (or a smaller
wavelength), the two approaches lead to consistent results as
expected.
\begin{figure}
\begin{center}
\includegraphics[scale=0.31]{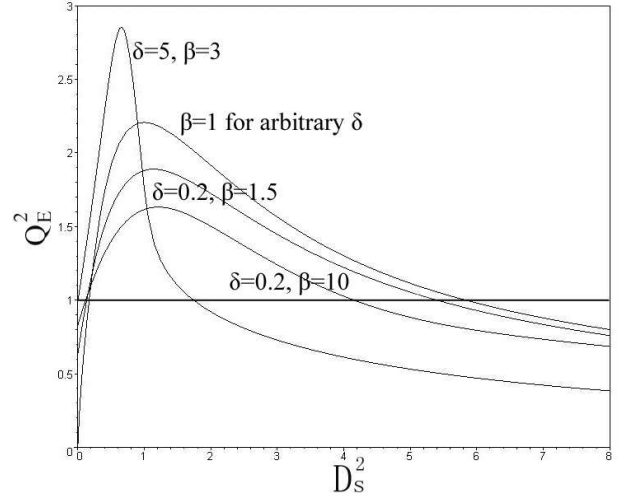}
\caption{Curves of $Q_E^2$ versus $D_s^2$ which intersect the
horizontal line $Q_E^2=1$. For each curve, the two points of
intersection at $Q_E^2=1$ give the stable range of $D_s^2$ for a
composite SID against axisymmetric disturbances. The $\beta=1$
case gives the same result of a single SID.}
\end{center}
\end{figure}

In our recent analysis (Lou \& Shen 2003), we obtained curves for
stationary axisymmetric perturbations given various combinations
of $\delta$ and $\beta$ for a composite SID system. These curves
differ from those for a single SID and the stable range of $D_s^2$
varies for different combinations of $\delta$ and $\beta$. It is
found that this stable range of $D_s^2$ may be significantly
reduced for certain values of $\delta$ and $\beta$, as shown in
Fig. $7-9$ of Lou \& Shen (2003) and in the analysis of Section
3.1. Moreover, when $\delta$ and $\beta$ are sufficiently large,
the stable range of $D_s^2$ may no longer exist, as shown in Fig.
10 of Lou \& Shen (2003) and Fig. 5 here. For these cases, a
composite SID system becomes vulnerable to axisymmetric
instabilities.

For $\delta=0.2$ and $\beta=10$, as shown in Fig. 6 of Lou \& Shen
(2003) and also Fig. 2 of this paper, we search for the
approximate stable range of $D_s^2$ with the WKBJ dispersion
relation (\ref{4}) for a composite SID system. By using the
$Q_E^2$ defined by equation (\ref{Qeff}), one can directly plot
the curve of $Q_E^2$ versus $D_s^2$ as shown in Fig. 6, from which
one can identify the approximate stable range of $D_s^2$ from
$\sim 0.1193$ to $\sim 4.1500$ where $Q_E^2=1$. Several other
curves with different $\delta$ and $\beta$ are also shown in Fig.
6.

One thing that should be emphasized is that the $Q_E^2$ parameter
as defined by equation (\ref{Qeff}) must be positive, otherwise it
may occasionally lead to incorrect conclusions regarding the sign
of $\omega_{-}^2$ as defined by equation (23). More detailed
investigations indicate that it is only applicable to cases where
a composite SID system can be stabilized in a proper range of
$D_s^2$. For special cases that cannot be stabilized at all, the
$Q_E^2$ thus defined may not be relevant.

\subsubsection{The $Q_J$ parameter of Jog}

The $Q_E$ parameter can be derived by solving cubic equation
(\ref{K3}) for $K$. It may happen that equation (\ref{K3}) gives
three real solutions. In order to obtain an analytical form of
$K_{min}$, one must identify the absolute minimum of
$\omega_{-}^2$. This procedure can be cumbersome. Alternatively,
Jog (1996) used a seminumerical way to define another effective
$Q$ parameter, referred to as $Q_J$ here, that is different from
$Q_E$ defined by equation (\ref{Qeff}).
\begin{figure}
\begin{center}
\includegraphics[scale=0.42]{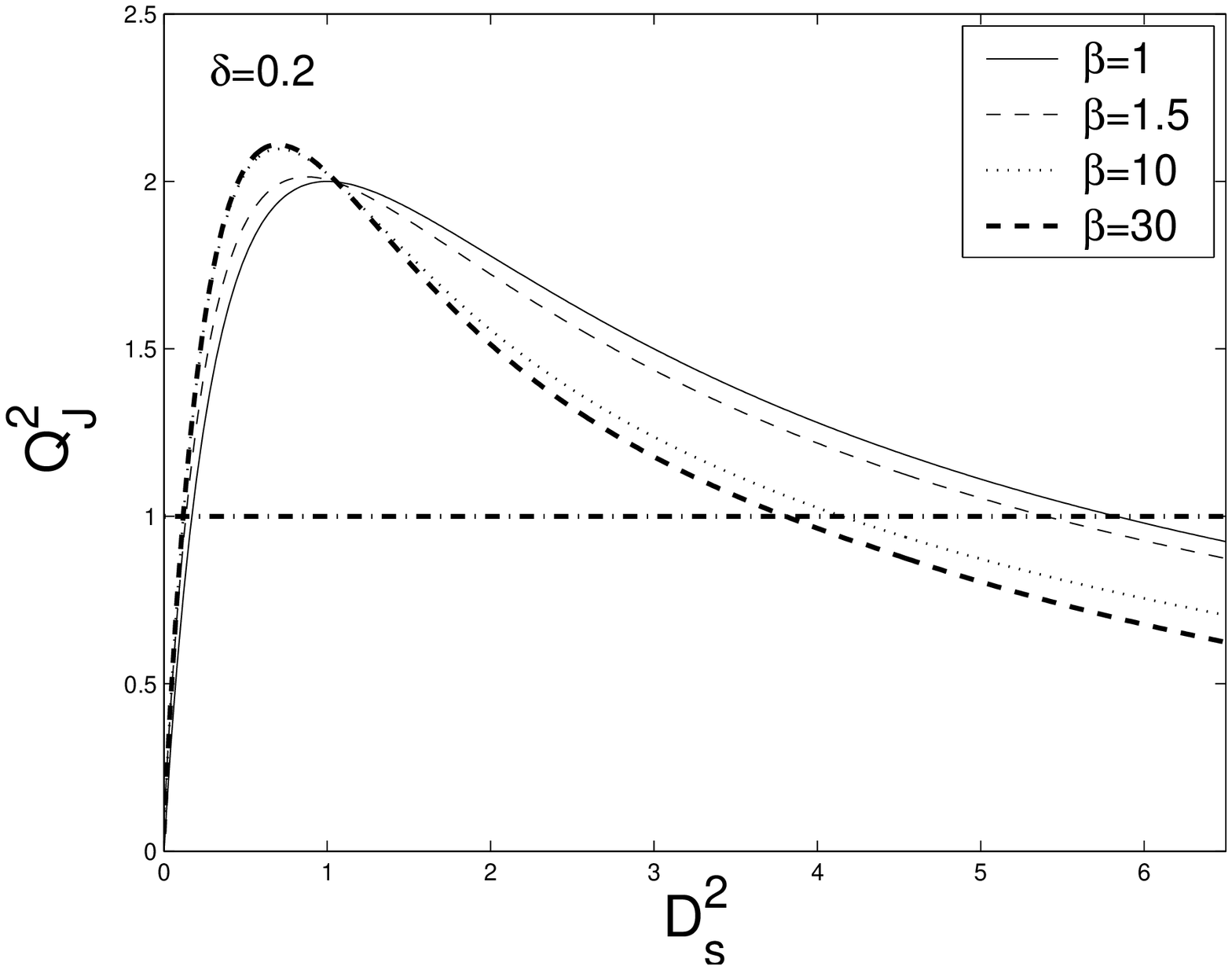}
\caption{Curves of $Q_J^2$ as function of $D_s^2$ with
$\delta=0.2$ but for different $\beta$ values. The stable ranges
of $D_s^2$ are determined by the two points of intersection with
the horizontal line $Q_J^2=1$. Stability corresponds to $Q_J^2>1$
when $D_s^2$ falls between the two intersection points. The two
intersection points move to the left as $\beta$ increases.}
\end{center}
%\end{figure}
%\begin{figure}[t]
\begin{center}
\includegraphics[scale=0.42]{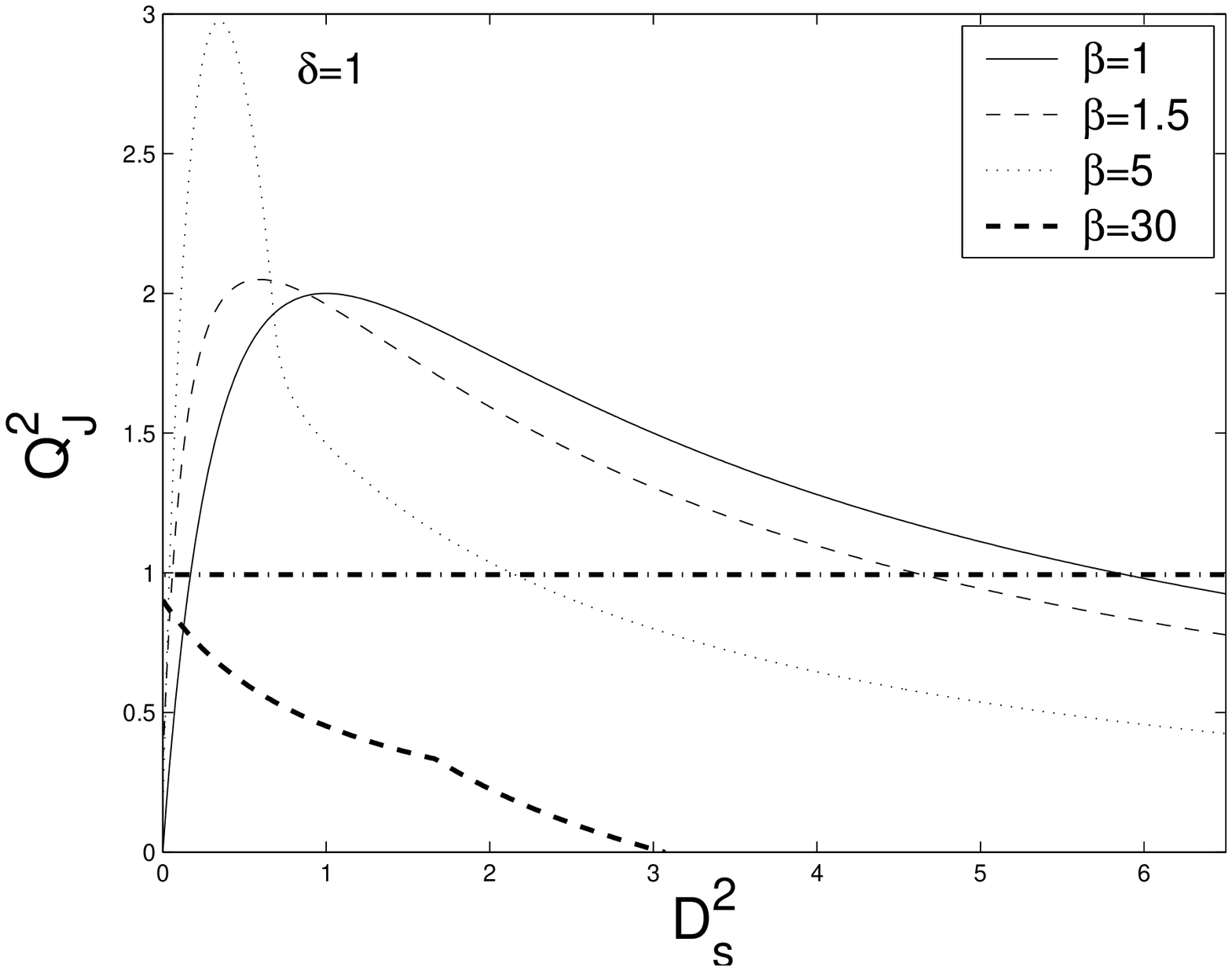}
\caption{Curves of $Q_J^2$ as function of $D_s^2$ with
$\delta=0.2$ but for different $\beta$ values. These curves
intersect with the horizontal line $Q_J^2=1$. For a larger
$\delta$ than that of Fig. 7, the two points of intersection
%the minimum and the maximum of $D_s^2$
move much faster as $\beta$ increases. And the stable range of
$D_s^2$ no longer exists when $\beta$ is sufficiently large. }
\end{center}
\end{figure}

According to solution (17), $\omega_{-}^2$ will be positive or
negative if $H_1H_2-G_1G_2>0$ or $<0$. The critical condition
of neutral stability is thus $H_1H_2-G_1G_2=0$, which can be
cast into the form of
$$
\frac{2\pi Gk\Sigma_0^s}{\kappa_s^2+k^2a_s^2}+\frac{2\pi
Gk\Sigma_0^g}{\kappa_g^2+k^2a_g^2}=1\ .
$$
One can now define a function ${\cal F}$ such that
\begin{equation}
\begin{split}
{\cal F}&\equiv\frac{2\pi Gk\Sigma_0^s}{\kappa_s^2+k^2a_s^2}
+\frac{2\pi Gk\Sigma_0^g}{\kappa_g^2+k^2a_g^2}\\
&=\frac{K(D_s^2+1)/(1+\delta)}{2D_s^2+K^2}+
\frac{K\beta(D_s^2+1)\delta/(1+\delta)}{2[\beta(D_s^2+1)-1]+K^2}\
,
\end{split}
\end{equation}
where expressions (2), (7), (8) and $K\equiv |k|r$ are used. We
then search for $K_{min}$ at which a composite SID system becomes
hardest to be stabilized, that is, when $\omega_{-}^2$ reaches the
minimum. The effective $Q$ parameter, $Q_J$, can thus be defined
as
\begin{equation}\label{QJ2}
\begin{split}
\frac{2}{1+Q_J^2}&\equiv
\frac{K_{min}(D_s^2+1)/(1+\delta)}{2D_s^2+K^2_{min}}\\
&\qquad +\frac{K_{min}\beta(D_s^2+1)\delta/(1+\delta)}
{2[\beta(D_s^2+1)-1]+K^2_{min}}\
\end{split}
\end{equation}
(see eqns. (5) and (6) of Jog 1996). It follows that $Q_J^2<1$
or $>1$ correspond to instability or stability, respectively.
Therefore, for a given set of $\delta$, $\beta$ and $D_s^2$, one
can numerically determine the value of $K_{min}$. By inserting
this $K_{min}$ into equation (\ref{QJ2}), one obtains the value of
$Q_J^2$ for any given set of $\{\delta,\beta,D_s^2\}$. It is then
possible to explore relevant parameter regimes of interest. Jog
(1996) introduced three parameters, namely, the two Toomre $Q$
parameters for each disc and the surface mass density ratio of the
two discs, to determine $Q_J^2$. Equivalently, we use three
dimensionless parameters $\delta$, $\beta$ and $D_s^2$ instead.
Figs. 7 and 8 show variations of $Q^2_J$ with $D_s^2$. Similar to
$Q_E^2$, the value of $Q_J^2>1$ when $D_s^2$ falls in the same
ranges. When $\delta$ is fixed and $\beta$ increases, the left and
right bounds of the stable $D_s^2$ range move towards left
together, while the width of the stable range appears to decrease.
We have revealed the same trend of variations in our recent work
(Lou \& Shen 2003). It is possible for the left bound to disappear
when $\beta$ is sufficiently large, and it is easier to achieve
this when $\delta$ becomes larger. For a sufficiently large $\delta$,
the increase of $\beta$ may completely suppress the stable range of
$D_s^2$ as shown in Fig. 5 for $\delta=1$ and $\beta=30$.

In comparison to $Q_E$, the $Q_J$ definition (\ref{QJ2}) is valid
for all parameter regimes and avoids improper situations for
unusual $\delta$ and $\beta$ values. However, to find the value of
$Q_J^2$ one must perform numerical exploration for each given
$D_s^2$, while for $Q_E^2$, the procedure is analytical as long as
there exists a stable range of $D_s^2$. Regardless, the critical
values of $D_s^2$ found by $Q_E^2$ and $Q_J^2$ are equivalent.

\subsection{Composite partial SID system}

In disc galaxies, there are overwhelming evidence for the existence
of massive dark matter halos as inferred from more or less flat
rotation curves. To mimic the gravitational effect of a
dark matter halo, we include gravity terms $\partial\Phi/\partial r$
and $\partial\Phi/\partial\varphi$ in the radial and azimuthal
momentum equations (4) and (5), respectively, where $\Phi$ is an
axisymmetric gravitational potential due to the dark matter halo.
It is convenient to introduce a dimensionless parameter
$F\equiv\phi/(\phi+\Phi)$ for the fraction of the disc
potential relative to the total potential (e.g., Syer \& Tremaine
1996; Shu et al. 2000; Lou 2002). The background rotational
equilibrium is thus modified by this additional $\Phi$ accordingly.
As before, we write
$\Omega_s=a_sD_s/r$, $\Omega_g=a_gD_g/r$, and
$\kappa_s=\sqrt{2}\Omega_s$, $\kappa_g=\sqrt{2}\Omega_g$. The
equilibrium surface mass densities now become
$$
\Sigma_0^s=\frac{a_s^2(1+D_s^2)F}{2\pi Gr(1+\delta)},
\qquad\qquad
\Sigma_0^g=\frac{a_g^2(1+D_g^2)F\delta}{2\pi Gr(1+\delta)}\ ,
$$
where $0\leq F\leq 1$. In our perturbation analysis, the
dynamic response of this axisymmetric massive dark matter
halo to coplanar perturbations in a composite SID system is
ignored\footnote{Strictly speaking, gravitational perturbation
coupling between a thin, less massive, rotating disc and a
more massive dark matter halo together with proper matching
and boundary conditions should lead to an increasing number
of modes that may or may not be stable. We simplify by
noting numerical simulations have indicated that a typical
galactic dark matter halo has a relatively high velocity
dispersion of the order of several hundred kilometers per
second. The rationale of the simplification is that
perturbation coupling between those in the SID system and
those in the dark matter halo becomes weak for a high velocity
dispersion in the dark matter halo. For example, for a non-rotating
spherical dark matter halo, effective ``acoustic" and ``gravity"
modes (i.e., $p-$modes and $g-$modes in the parlance of stellar
oscillations; Lou 1995) can exist within; these $p-$ and $g-$modes
are expected to be Landau damped in a collisionless system.
However, the stability of ``interface modes" between a thin
rotating disc and a dark matter halo remains to be investigated. }
(Syer \& Tremaine 1996; Shu et al. 2000; Lou 2002).
We thus derive similar perturbation equations as those of a
full SID case ($F=1$) and consequently the similar dispersion
relation as equation (13) yet with modified equilibrium properties.
\begin{figure}
\begin{center}
\includegraphics[scale=0.41]{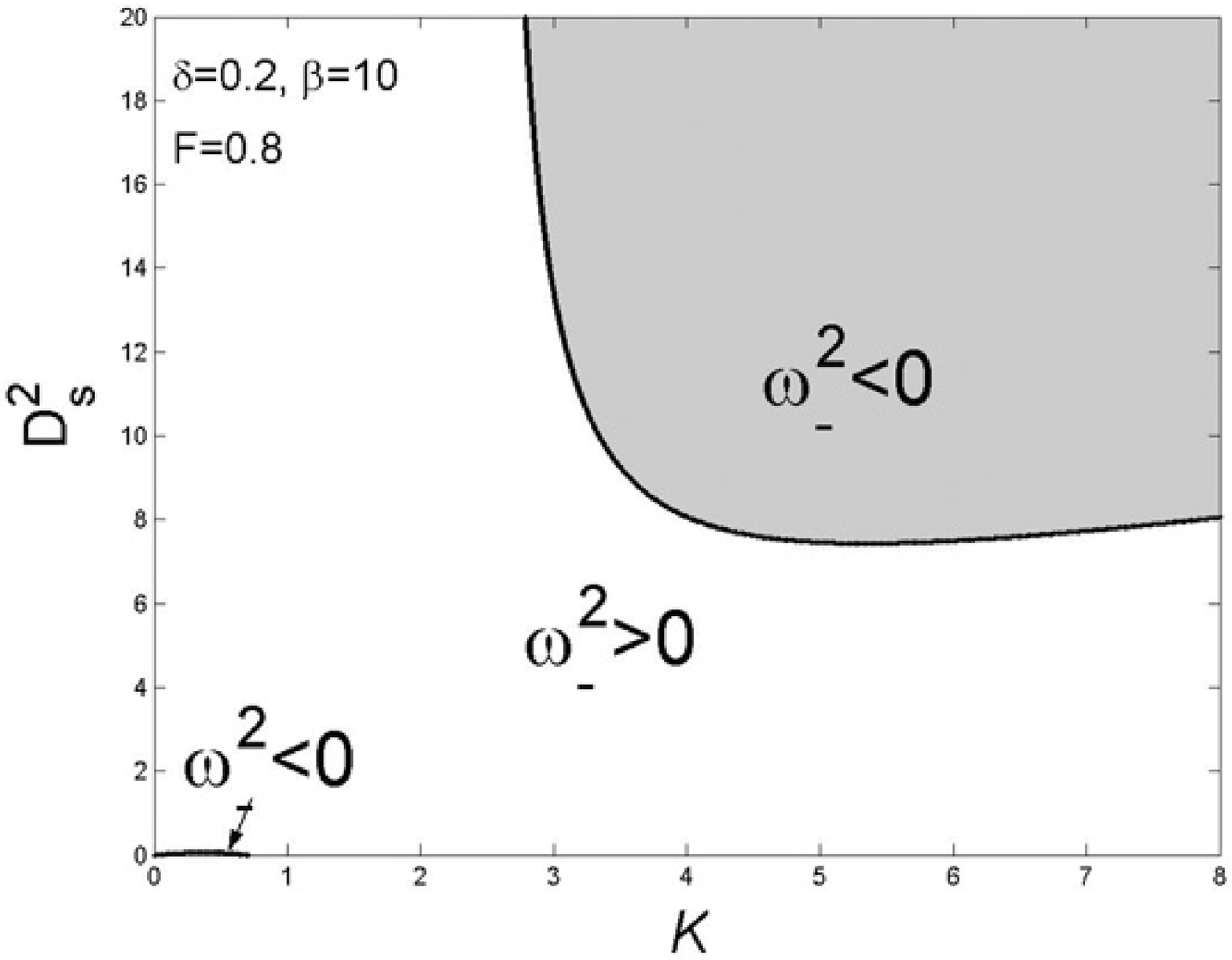}
\caption{Contours of $\omega_{-}^2$ as function of stellar
rotation parameter $D_s^2$ and wavenumber $K$ for a composite
partial SID system with $\delta=0.2$, $\beta=10$ and $F=0.8$. }
\end{center}
%\end{figure}
%\begin{figure}
%\begin{center}
%\includegraphics[scale=0.41]{delta_0.2beta_10F_0.5.eps}
%\caption{Contours of $\omega_{-}^2$ as a function of stellar
%rotation parameter $D_s^2$ and wavenumber $K$ for a composite
%partial SID system with $\delta=0.2$, $\beta=10$ and $F=0.5$. }
%\end{center}
%\end{figure}
%\begin{figure}[t]
\begin{center}
\includegraphics[scale=0.41]{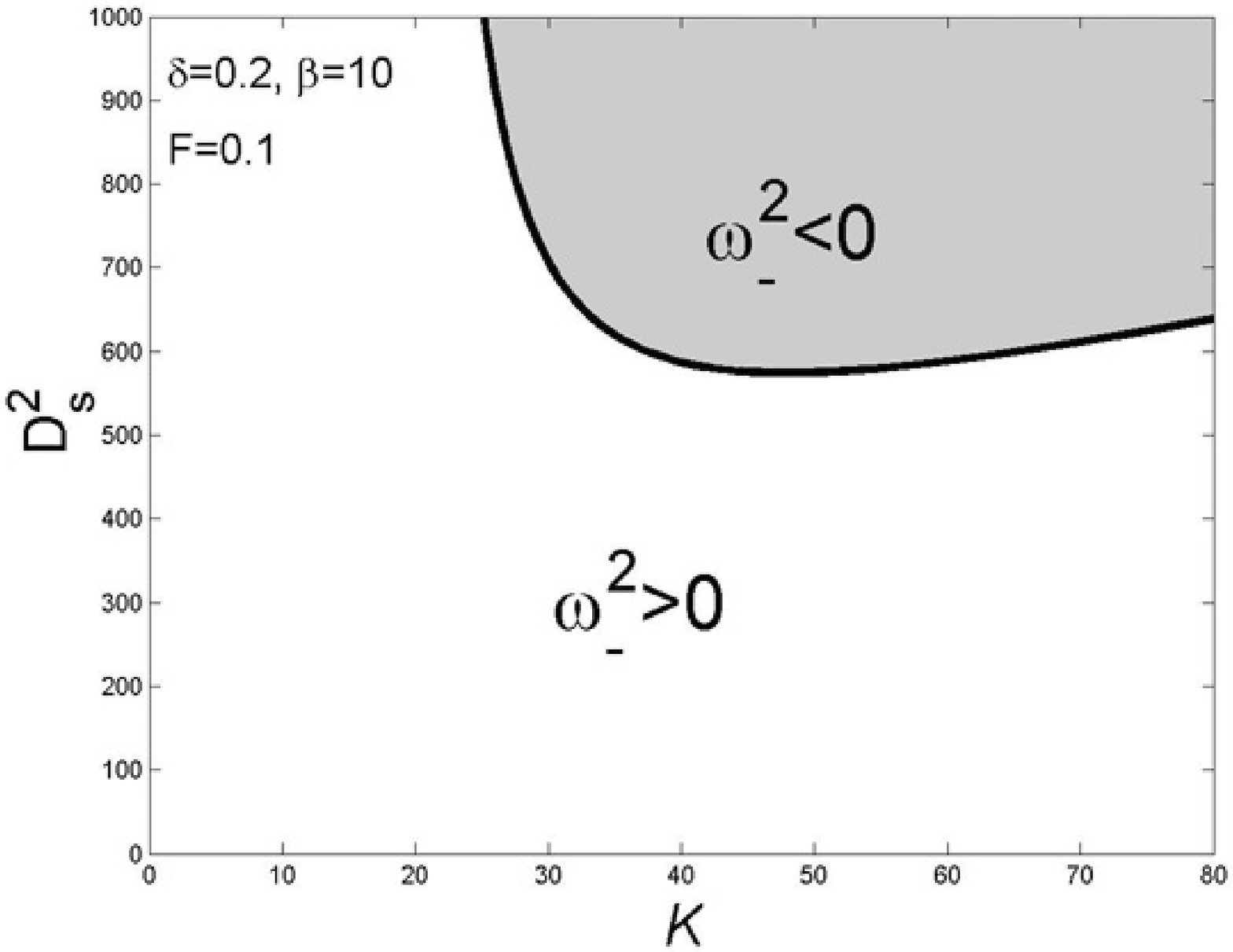}
\caption{Contours of $\omega_{-}^2$ as function of stellar
rotation parameter $D_s^2$ and wavenumber $K$ for a composite
partial SID system with $\delta=0.2$, $\beta=10$ and $F=0.1$. }
\end{center}
\end{figure}
Now following the same procedure of WKBJ perturbation analysis
described in Section 3.1, we can plot contours of $\omega_{-}^2$
as a function of rotation parameter $D_s^2$ and wavenumber $K$ for
a composite partial SID system. The case of $F=1$ corresponds to a
composite full SID system that has been studied in Section 3. When
$F$ becomes less than 1 (i.e., a composite partial SID system),
which means the fraction of the dark matter halo increases, the
stable range of $D_s^2$ becomes enlarged, as shown in Figs. 9 and
10 for $\delta=0.2$ and $\beta=10$. From these contour plots, it
is clear that the introduction of a dark matter halo tends to
stabilize a composite partial SID system. For late-type disc
galaxies, one may take $F=0.1$ or smaller. Such composite partial
SID systems are stable against axisymmetric disturbances in a wide
range of $D_s^2$. Moreover, those composite full SID systems that
are unstable may be stabilized by the presence of a dark matter
halo, as shown by the example of Fig. 11 for $\delta=1$, $\beta=30$
and $F=0.5$ (this case is unstable for a composite full SID system).
\begin{figure}
\begin{center}
\includegraphics[scale=0.41]{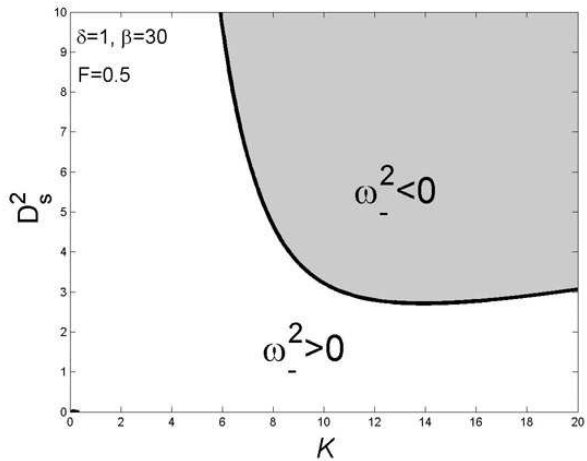}
\caption{Contours of $\omega_{-}^2$ as function of stellar
rotation parameter $D_s^2$ and wavenumber $K$ for a composite
partial SID system with $\delta=1$, $\beta=30$ and $F=0.5$. }
\end{center}
\end{figure}

\section{Discussions and summary}

Our model of a composite (full or partial) SID system
is an idealization of any actual disc systems, such as
disc galaxies, circumnuclear discs surrounding nuclei of
galaxies, or protostellar discs.
One key feature of the SID model is the flat rotation curve.
One major deficiency in galactic application is the central
singularity that sometimes even gives rise to conflicting
theoretical interpretations (see Shu et al. 2000). The
isothermality is yet another simplifying approximation. In
galactic applications, the common wisdom is to introduce a
bulge around the center to avoid the central singularity.
By a local WKBJ stability analysis of axisymmetric perturbations,
this model can be utilized as usual with qualifications. We now
discuss the WKBJ results in the context of our own Galaxy, the
Milky Way, for which the necessary observational data are available,
such as the gas fraction, the stellar and the gaseous velocity
dispersions and the epicyclic frequency, and so forth.

There are several inadequacies of our model. First, the isothermal
assumption implying constant $a_s$ and $a_g$ throughout the disc
system is a gross simplification; the velocity dispersion $a_s$ in
the Milky Way decreases with increasing radius (e.g., Lewis \&
Freeman 1989). Secondly, by the polytropic approximation, the
surface mass densities are characterized by power-law $\propto
r^{-1}$ profiles, which appear not to be the case of the Milky Way
(e.g., Caldwell \& Ostriker 1981). Nevertheless, to estimate local
stability properties of a disc portion, we may take $a_s=50\hbox{
km s}^{-1}$, $V_s=220\hbox{ km s}^{-1}$, $\delta=0.1$ and flat
rotation speed $V_s=220\hbox{ km s}^{-1}$ around $4\hbox{ kpc}$
from the center of our Galaxy. For a massive dark matter halo such
that $F=0.1$, the relevant parameters to determine the effective
$Q$ parameter for a composite partial SID system are $\delta=0.1$,
$\beta=50$, $D_s^2=20$ and $F=0.1$. Following the definition of
Jog (1996), the value of $Q_J\gg1$. While for full SIDs with
$F=1$, we get $Q_J=0.17$ which means local instability. Obviously,
the dark matter halo plays the key role in stabilizing the system.

The composite partial SID model with power-law $\propto r^{-1}$
surface mass densities may not describe the Milky Way well. But
for those disc galaxies with flat rotation curves and approximate
power-law $\propto r^{-1}$ surface mass densities, our model
should offer a sensible local criterion for axisymmetric
instability. For example, given radial variations of velocity
dispersions and gas mass fraction, we may deduce locally unstable
zones.

For circumnuclear discs around nuclei of galaxies, the stellar
velocity dispersion $a_s$ can be as high as several hundred
kilometers per second along the line of sight (e.g., Whitmore
et al. 1985; McElroy 1994); this may then lead to small $D_s^2$.
Within such a parameter regime, the collapse instability may set
in during a certain phase of system evolution. For instance, in
the case of Fig. 2 with $\delta=0.2$ and $\beta=10$, we choose
$V_s=200\hbox{ km s}^{-1}$ and $a_s=300\hbox{ km s}^{-1}$ in a
circumnuclear disc region of a disc galaxy. The resulting
$D_s^2=0.44$ would give rise to collapse and the composite SID
system might eventually evolve into a bulge of high velocity
dispersion.

Another possible collapse situation of interest corresponds
to small $D_s^2$ for sufficiently low disc rotation speed $V$.
Potential applications include but not limited to protostellar
discs in the context of star formation. We here briefly discuss
several qualitative aspects. For a protostellar disc system, the
two disc components here may be identified with the relatively
``hot" gas disc and the relatively ``cool" dust disc.
Interactions of radiation fields from the central protostar
with cool dusts in the disc lead to infrared emissions.
In our idealized treatment, the gas and dust discs
are treated as two gravitationally coupled thin SIDs.
This simple model treatment may need to be complemented
by other processes of coupling such as
collisions between gas and dust particles.
When such a composite disc system rotates at a low speed
(e.g., Shu et al. 1987), the rotation parameter $D$ could
be low enough to initiate large-scale collapse.
The subsequent dynamical processes will drive the composite
SID system to more violent star formation activities.

So far we have investigated the axisymmetric stability problem for
a composite system of gravitationally coupled stellar and gaseous
SIDs using the WKBJ approximation and we now summarize the results
of our analysis.

First, because the single SID case studied by Shu et al. (2000)
may be regarded as the special case of a composite SID system,
the results of our WKBJ analysis clearly support the physical
interpretations of Shu et al. (2000) for the marginal stability
curves of axisymmetric perturbations. The ``ring fragmentation
regime" is related to the familiar $Q$ parameter (Safronov 1960;
Toomre 1964). The presence of the Jeans ``collapse regime" can be
traced to the SID model where self-gravity, effective pressure,
and SID rotation compete with each other.

Secondly, the recent study of Lou \& Shen (2003) generalizes that
of Shu et al. (2000) and describes exact perturbation solutions in
a composite SID system. We obtained marginal stability curves for
axisymmetric perturbations that are qualitatively similar to those
of Shu et al. (2000) and that depend on additional dimensionless
parameters. It is natural to extend the interpretations of Shu et
al. (2000). Through the WKBJ analysis here, we now firmly
establish the presence of two regimes of ``ring fragmentation" and
``collapse" in a composite SID system. It is fairly straightforward
to apply our exact $D-$criterion for a composite SID system, not
only for disc galaxies, but also for other disc systems including
circumnuclear discs and protostellar discs etc.

Thirdly, in the WKBJ approximation, it is also possible to relate
our $D-$criterion to the $Q_E$ criterion similar to that of
Elmegreen (1995). Because $\kappa_s$ and $\kappa_g$ are different
in general, our WKBJ treatment is not the same as that of
Elmegreen (1995).

Fourthly, also in the WKBJ approximation, we relate our
$D-$criterion to the $Q_J$ criterion as defined by Jog (1996) but
with $\kappa_s\neq\kappa_g$ in general. We find that the $Q_J$
criterion is fairly robust in the WKBJ regime.

Finally, we further consider the axisymmetric stability of a
composite partial SID system to include the gravitational effect
from an axisymmetric dark matter halo. The stabilizing effect of
the dark matter halo is apparent.

\section*{Acknowledgments}
We thank the referee C. J. Jog for comments and suggestions.
This research has been supported in part by the ASCI Center for
Astrophysical Thermonuclear Flashes at the University of Chicago
under Department of Energy contract B341495, by the Special Funds
for Major State Basic Science Research Projects of China, by the
Tsinghua Center for Astrophysics, by the Collaborative Research
Fund from the NSF
%National Natural Science Foundation
of China (NSFC) for Young Outstanding Overseas Chinese Scholars
(NSFC 10028306) at the National Astronomical Observatory, Chinese
Academy of Sciences, and by the Yangtze Endowment from the
Ministry of Education through the Tsinghua University. Affiliated
institutions of Y.Q.L. share the contribution.

%\appendix
%\section\\
%We now prove in most cases equation (\ref{K3}) only yields one
%real solution about $K$. The determinant of the cubic equation
%about $K$ is
%\begin{equation}
%\begin{split}
%\Delta&=q^2/4+p^3/27\\
%\end{split}
%\end{equation}

\end{document}